\documentclass[RevStyle,iicol]{paper}

\usepackage{graphicx}

\begin{document}

\title[EW corrections to inclusive $B$ decays]{Finite $W$ mass corrections to inclusive $B$ decays}

\author[]{\fnm{Dante} \sur{Bigi}}\email{bigid@iol.it}

\author[1,2]{\fnm{Sandro} \sur{Mächler}}\email{sandro.maechler@physik.uzh.ch}

\affil[1]{\orgdiv{Dipartimento di Fisica}, \orgname{Università di Torino \& INFN, Sezione di Torino}, \orgaddress{\street{Via Pietro Giuria}, \postcode{I-10125}, \city{Turin}, \country{Italy}}}

\affil[2]{\orgdiv{Physikinstitut}, \orgname{Universität Zürich}, \orgaddress{\street{Winterthurerstrasse 190}, \postcode{CH-8057}, \city{Zürich}, \country{Switzerland}}}

\abstract{We compute the effects of a finite $W$ boson mass on the inclusive $b\rightarrow c \ell \nu$ decay rate and the semileptonic moments. To this end we keep terms of $\mathcal{O}\left(q^2/m_W^2 \right)$ when integrating out the $W$ boson in the Weak Effective Theory. In the third hadronic invariant mass moment such corrections reach the $0.5\%$ level but they are smaller than the present-day experimental uncertainties in all observables.}

\maketitle
\section{Introduction}\label{sec1}
Over recent years the study of semileptonic decays of $B$ mesons has been developing rapidly, with intriguing tensions between Standard Model (SM) predictions and experimental measurements appearing. One of the most persistent tensions in this field is the $\sim 3 \sigma$ tension between the exclusive and inclusive determinations of the CKM matrix element $V_{cb}$\cite{Workman:2022ynf,HFLAV:2019otj,Gambino:2019sif,Gambino:2020jvv}.

On the inclusive side the theoretical description relies on the Heavy Quark Expansion (HQE) \cite{Chay:1990da,Manohar:1993qn,Blok:1993va}, an expansion in powers of $\Lambda_\mathrm{QCD}/m_b$. By now the HQE is at a mature state with the corrections up to $\mathcal{O}\left(\Lambda_\mathrm{QCD}^5/m_b^5 \right)$ known at tree-level \cite{Bigi:2009ym,Mannel:2010wj,Gambino:2016jkc}, the one--loop correction fully known at $\mathcal{O}\left(\Lambda_\mathrm{QCD}^2/m_b^2 \right)$ \cite{Becher:2007tk,Alberti:2012dn,Alberti:2013kxa,Mannel:2014xza,Mannel:2015jka} and in the case of the lepton invariant mass spectrum even at $\mathcal{O}\left(\Lambda_\mathrm{QCD}^3/m_b^3 \right)$ \cite{Mannel:2021zzr}. At the leading order in the HQE the three--loop corrections to the total decay rate and the semileptonic moments have been computed recently \cite{Fael:2020tow,Fael:2022frj}.

The starting point of any HQE computation is the Weak Effective Theory (WET) with the $W$ boson integrated out at the electroweak scale. The leading corrections to the limit of an infinitely heavy $W$ boson are given by terms of $\mathcal{O}(q^2/m_W^2)$, where $q^2$ is the dilepton invariant mass. Being suppressed by the $W$ boson mass, these corrections are expected to be small. Nonetheless, in principle they could have a sizeable effect on the moments of the dilepton invariant mass spectrum recently used to determine the CKM matrix element $V_{cb}$ \cite{Bernlochner:2022ucr}. In conjunction with the current theory precision this renders a computation of $\mathcal{O}\left(q^2/m_W^2 \right)$ effects on the inclusive decay rate and moments timely. The purpose of this paper is the presentation of such a computation.

\section{Triple differential decay width to \texorpdfstring{$\mathcal{O}(q^2/m_W^2) $}{}}\label{sec2}
We consider the inclusive decay $ B(p) \rightarrow X_c(p-q) \ell(p_\ell) \nu_\ell(p_\nu) $ of a $B$ meson into a charmed hadronic final state and lepton pair.
As usual we start by integrating the $W$ boson out of our description physics at the $B$ scale. We do however keep terms of $\mathcal{O}(q^2/m_W^2)$ in the expansion of the $W$ propagator, leading to the effective Hamiltonian
\begin{align} \label{eq:EffectiveHamiltonian}
    \mathcal{H}_\mathrm{eff} = \frac{G_\mathrm{F}}{\sqrt{2}}V_{cb} &\Big [g_{\mu\nu} \left(1+\frac{D^2}{m_W^2} \right) J_q^\mu J_\ell^\nu  \nonumber \\
    &- \frac{D_\mu D_\nu}{m_W^2} J_q^\mu J_\ell^\nu \Big],
\end{align}
 where $J_q^\mu(x) = \left(\overline{c}\gamma^\mu (1-\gamma_5) b \right)$ and $J_\ell^\mu(x) = \left(\overline{\ell} \gamma^\nu (1-\gamma_5) \nu_\ell \right)$ are the hadronic and leptonic currents and the covariant derivative is given by
\begin{equation}
    D_\mu = \partial_\mu + ig T^a G_\mu^a + ie Q A_\mu,
\end{equation}
with the $\mathrm{SU}(3)_c$ and $\mathrm{U}(1)_Q$ gauge fields $G_\mu^a$ and $A_\mu$. In the following we consider light leptons ($\ell \in \{e, \mu\} $). In this limit the last term of Eq. (\ref{eq:EffectiveHamiltonian}) does not contribute.

Now we can write the triple differential decay rate as
\begin{align}
    &\frac{8\pi^3}{G_\mathrm{F}^2\left\vert V_{cb} \right\vert^2}\frac{\mathrm{d}\Gamma}{\mathrm{d}q^2 \mathrm{d}q^0 \mathrm{d}E_\ell} \nonumber \\
    &=L_{\mu \nu} g^{\mu\alpha}\left(1+\frac{q^2}{m_W^2} \right) g^{\nu\beta}\left(1+\frac{q^2}{m_W^2} \right) W_{\alpha\beta} \nonumber \\
    &=\left(1+\frac{2q^2}{m_W^2} \right) L_{\mu \nu}W^{\mu\nu} + \mathcal{O}\left( \frac{q^4}{m_W^4} \right).
\end{align}
Here 
\begin{equation}
    L^{\mu\nu}=p_\ell^\mu p_\nu^\nu -p_\ell \cdot p_\nu g^{\mu\nu} + p_\ell^\nu p_\nu^\mu + i\epsilon^{\mu\alpha\nu\beta}p_{\ell,\alpha} p_{\nu,\beta}
\end{equation}
is the standard leptonic tensor and the hadronic tensor is defined as
\begin{align}
    &\frac{2m_B}{(2\pi)^3}W^{\mu\nu} = \nonumber \\
    &\sum_{X_c} \delta^{(4)}(p_B-q-p_X) \left\langle B \left\vert J_q^{\mu\dagger} \right\vert X_c \right\rangle \left\langle X_c \left\vert J_q^\nu \right\vert B \right\rangle.
\end{align}
The hadronic tensor can be decomposed into Lorentz invariant structure functions as
\begin{equation}
    W^{\mu\nu} = -g^{\mu\nu}W_1 + v^\mu v^\nu W_2 -i\epsilon^{\mu\nu\alpha\beta}v_\alpha\hat q_\beta W_3,
\end{equation}
where $v = p/m_B$ is the $B$ meson four velocity and $\hat q = q/m_b$. Then in the rest frame of the decaying B-meson we have $v\cdot p_\ell = E_\ell$ and $v\cdot q = q_0$. Consequently
\begin{align}
    L_{\mu\nu} W^{\mu\nu} = & q^2 W_1 +\left(2E_\ell q_0 - 2 E_\ell^2 -\frac{q^2}{2} \right)W_2 \nonumber \\
    &+ q^2\left(2E_\ell - q_0 \right) W_3.
\end{align}
The structure functions are computed in the HQE, resulting in a double expansion in terms of inverse powers of the $b$ quark mass and powers of the strong coupling constant. The current state of the art of the HQE is summarized in Table \ref{Tab:AvailableCorrections}.

\section{Results}
We study the impact of $\mathcal{O}(q^2/m_W^2)$ contributions on the total decay rate and the moments of the charged lepton energy $E_\ell$, the hadronic invariant mass $m_X$ and the invariant mass of the lepton pair $q^2$. The moments are defined as
\begin{equation}\label{Eq:MomentDefinition}
    \left\langle x^n \right\rangle = \frac{1}{\Gamma}\int_{E_\ell > E_\ell^{\mathrm{cut}}, \; q^2 > q^2_\mathrm{cut}} \mathrm{d}q^2 \mathrm{d}q_0 \mathrm{d}E_\ell x^n \frac{\mathrm{d}\Gamma}{\mathrm{d}q^2 \mathrm{d}q_0 \mathrm{d}E_\ell}
\end{equation}
with
\begin{equation}
    \Gamma = \int_{E_\ell > E_\ell^{\mathrm{cut}}, \; q^2 > q^2_\mathrm{cut}} \mathrm{d}q^2 \mathrm{d}q_0 \mathrm{d}E_\ell \frac{\mathrm{d}\Gamma}{\mathrm{d}q^2 \mathrm{d}q_0 \mathrm{d}E_\ell}
\end{equation}
for $x\in \{E_\ell, m_X, q^2 \}$. In the leptonic and hadronic moments a lower cut on the energy of the charged lepton, $E_\ell$, is applied while in the case of the $q^2$ a lower cut on the invariant mass of the lepton pair is applied. For the lepton energy and $q^2$ moments these integrals can be evaluated directly. In the case of the hadronic mass moments however we compute the moments of $\hat u = ((p - q)^2 - m_c^2)/m_b^2$ and $\hat \omega = 1 - q_0/m_b$ and relate them to the hadronic final state mass through
\begin{equation}
    \left\langle m_X^2 \right\rangle = m_c^2+\Lambda^2 +2\Lambda m_b \left\langle \hat \omega \right\rangle + m_b^2 \left\langle \hat u \right\rangle,
\end{equation}
where $\Lambda = m_B - m_b$ is the mass difference between the $B$ meson and the $b$ quark. For $n>1$ one usually considers the central moments $\left\langle \left(x - \left\langle x \right\rangle \right)^n \right\rangle$. We denote them as
\begin{align}
    \ell_i &= \begin{cases} \left\langle E_\ell \right\rangle, \text{ for } i=1 \\ \left\langle \left(E_\ell - \left\langle E_\ell \right\rangle \right)^i \right\rangle, \text{ for } i>1 \end{cases} \nonumber\\
    h_i &= \begin{cases} \left\langle m_X^2 \right\rangle, \text{ for } i=1 \\ \left\langle \left(m_X^2 - \left\langle m_X^2 \right\rangle \right)^i \right\rangle, \text{ for } i>1 \end{cases} \nonumber\\
    \mathcal{Q}_i &= \begin{cases} \left\langle q^2 \right\rangle, \text{ for } i=1 \\ \left\langle \left(q^2 - \left\langle q^2 \right\rangle \right)^i \right\rangle, \text{ for } i>1. \end{cases}
\end{align}
By carrying out the integrations in Eq. \ref{Eq:MomentDefinition} and re-expanding the ratio in powers of $1/m_b$, $\alpha_s$ and $\xi = 2m_b^2/m_W^2$ we find the contributions of $\mathcal{O}(q^2/m_W^2)$. As the analytic expressions are lengthy they are attached to the arXiv submission in an ancillary file. Numerically we find the results in Tables \ref{tab:LeptonMomentsq2mW}-\ref{tab:q2Momentsq2mW}.

\begin{table}[!ht]
    \centering
    \begin{tabular}{c c}
        Parameter & Value \\
        \hline
        $m_b^\mathrm{kin}(m_b)$ & $4.6 \text{ GeV}$ \\
        $\overline{m}_c(2\mathrm{GeV})$ & $1.15 \text{ GeV}$ \\
        $\mu_\pi^2$ & $0.4 \text{ GeV}^2$ \\
        $\mu_G^2$ & $0.35 \text{ GeV}^2$ \\
        $\rho_D^3$ & $0.2 \text{ GeV}^3$ \\
        $\rho_{LS}^3$ & $-0.15 \text{ GeV}^3 $
    \end{tabular}
    \caption{Default input values}
    \label{tab:Default_inputs}
\end{table}

\begin{table}[!ht]
    \centering
    \begin{tabular}{c c c c c}
         & \multicolumn{3}{c}{$10^4 \times \left.\mathcal{O}\left(\frac{q^2}{m_W^2} \right)\right/\mathrm{LO}$} \\
         \hline
         $E_\mathrm{cut}$ & $\Gamma$ & $\ell_1$ & $\ell_2$ & $\ell_3$ \\
         \hline
         0 & $14.3$ & $1.134$ & $-3.140$ & $3.212$ \\
         1 GeV & $15.9$ & $0.368$ & $-1.53043 $ & $25.4 $
    \end{tabular}
    \caption{$\mathcal{O}(q^2/m_W^2))$ contributions to the total decay rate and leptonic moments }
    \label{tab:LeptonMomentsq2mW}
\end{table}

\begin{table}[!ht]
    \centering
    \begin{tabular}{c c c c}
         & \multicolumn{3}{c}{$10^4 \times \left.\mathcal{O}\left(\frac{q^2}{m_W^2} \right)\right/\mathrm{LO}$} \\
         \hline
         $E_\mathrm{cut}$ & $h_1$ & $h_2$ & $h_3$ \\
         \hline
         0 & $-0.933 $ & $3.50 $ & $-21.0 $ \\
         1 GeV & $-0.946 $ & $1.77 $ & $-46.2 $
    \end{tabular}
    \caption{$\mathcal{O}(q^2/m_W^2))$ contributions to the hadronic moments }
    \label{tab:HadronMomentsq2mW}
\end{table}

\begin{table}[!ht]
    \centering
    \begin{tabular}{c c c c}
         & \multicolumn{3}{c}{$10^4 \times \left.\mathcal{O}\left(\frac{q^2}{m_W^2} \right)\right/\mathrm{LO}$} \\
         \hline
         $q^2_\mathrm{cut}$ & $\mathcal{Q}_1$ & $\mathcal{Q}_2$ & $\mathcal{Q}_3$ \\
         \hline
         0 & $6.04 $ & $3.50 $ & $-21.0 $ \\
         3 $\text{GeV}^2$ & $2.47 $ & $2.87 $ & $-14.2 $
    \end{tabular}
    \caption{$\mathcal{O}(q^2/m_W^2))$ contributions to the leptonic invariant mass moments }
    \label{tab:q2Momentsq2mW}
\end{table}

In Appendix \ref{sec:CutDependence} the full dependence of the moments on the experimental cuts is illustrated by the examples of the first moments of the charged lepton energy, the hadronic invariant mass and the dilepton invariant mass. The charged lepton energy moments have been measured by the DELPHI \cite{DELPHI:2005mot}, BaBar \cite{BaBar:2004bij} and Belle \cite{Belle:2006kgy} collaborations, the hadronic invariant mass moments have been measured by the BaBar \cite{BaBar:2007dwo,BaBar:2009zpz}, Belle \cite{Belle:2006jtu}, Belle II \cite{Belle-II:2020oxx}, CDF \cite{CDF:2005xlh}, CLEO \cite{CLEO:2004bqt} and DELPHI \cite{DELPHI:2005mot} collaborations and the $q^2$ moments have been measured by the Belle \cite{Belle:2021idw}, Belle II \cite{Belle-II:2022fug} and CLEO \cite{CLEO:2004bqt} collaborations. The Belle II uncertainties are not plotted because they are larger than the ones of the older measurements. As can be seen in the plots the $q^2/m_W^2$ corrections decrease for higher cuts and they are multiple orders of magnitude smaller than the experimental uncertainties. Therefore they are unlikely to significantly impact the theory prediction of $V_{cb}$.

They are however of the same order of magnitude as the $\mathcal{O}(\alpha_s^3)$ for the total rate, the second central lepton energy moment and the first hadronic mass moment. Therefore they may also be considered if one keeps the three--loop corrections, in order to consistently treat contributions of their order of magnitude.

\section{Conclusions}
We computed the corrections of $\mathcal{O}\left(q^2/m_W^2 \right)$ to the inclusive total rate and moments in semileptonic $B$ decays. In most kinematical distributions these effects are several orders of magnitude smaller than the current experimental uncertainties. However, they are of comparable size to the three--loop corrections, which are included in present analyses. Thanks to the results presented in this paper, they can be consistently included in future studies.

\bmhead{Acknowledgments}
DB and SM would like to thank Paolo Gambino for discussions and proposing the project. SM is grateful to Gino Isidori, Zach Polonksy and Jason Aebischer for useful discussions. This project has received funding from the Swiss National Science Foundation (SNF) under contract 200020\_204428.
\begin{appendices}

\section{Currently available corrections}
In Table \ref{Tab:AvailableCorrections} we list the currently available corrections and where to find them. Note that in different references different input values for the quark masses, HQE matrix elements, the strong coupling constant and the kinematic cut are used so these numbers can not be directly compared to each other. Nonetheless, their respective orders of magnitude are informative and having them all collected in one place might be useful to someone who is not yet familiar with the HQE.

The values in the first two columns, i.e. the corrections of $\mathcal{O}\left(\Lambda_\mathrm{QCD}^2 / m_b^2 \right) $ and $\mathcal{O}\left(\Lambda_\mathrm{QCD}^3 / m_b^3 \right) $ are computed from the triple differential decay rates given in \cite{Blok:1993va} and \cite{Gremm:1996df} respectively.

The $\mathcal{O}\left(\Lambda_\mathrm{QCD}^2 / m_b^2 \right) $ and $\mathcal{O}\left(\Lambda_\mathrm{QCD}^3 / m_b^3 \right) $ corrections to the central moments of $q^2$ are computed from the expressions of the ancillary file supplied with \cite{Fael:2018vsp} using the default input values given in Table \ref{tab:Default_inputs} with a cut of $q^2_\mathrm{cut} = 3 \mathrm{GeV}^2$. While for the $\mathcal{O}\left(\Lambda_\mathrm{QCD}^4 / m_b^4 \right) $ the results of the results of the fit including all $1/m_b^4$ parameters in \cite{Bernlochner:2022ucr} are used for the input parameters and a cut of $q_\mathrm{cut} = 3 \mathrm{GeV}^2$ is applied.
\begin{sidewaystable*}[!ht]
    \caption{Available contributions. The input parameters can be found in the respective references. In the radiative corrections the reference values in the kinetic scheme with a cutoff of $\mu_\mathrm{kin} = 1 \mathrm{GeV}$ and charm mass defined in the $\overline{\mathrm{MS}}$ scheme evaluated at $2 \text{ GeV}$ are cited.}
    \label{Tab:AvailableCorrections}
    \resizebox{\textheight}{!}{%
    \begin{tabular}{l|c|c|c|c|c|c|c|c|c|c|c}
         & $\frac{\Lambda_{\mathrm{QCD}}^2}{m_b^2}$ & $\frac{\Lambda_{\mathrm{QCD}}^3}{m_b^3}$ & $\frac{\Lambda_{\mathrm{QCD}}^{4}}{m_b^{4}}$ & $\frac{\Lambda_{\mathrm{QCD}}^{5}}{m_b^{5}}$ & $\alpha_s$ & $\beta_0 \alpha_s^2 $ & $\alpha_s^2$ & $\alpha_s^3$ & $\alpha_s \frac{\Lambda_{\mathrm{QCD}}^2}{m_b^2}$ & $\alpha_s \frac{\Lambda_{\mathrm{QCD}}^3}{m_b^3}$ & $\frac{q^2}{m_W^2}$ \\
         \hline
         $\mathrm{d}\Gamma/\mathrm{d}x^3$ & \cite{Blok:1993va} & \cite{Gremm:1996df} & &  & \cite{Aquila:2005hq} & \cite{Aquila:2005hq} & & \cite{Fael:2020tow} & \cite{Alberti:2013kxa} & & \\
         \hline
        $\Gamma$ & $-4\%$ & $-3\%$ & $\mathcal{O}(1\%) $ \cite{Mannel:2010wj} & $\mathcal{O}(0.5\%) $ \cite{Mannel:2010wj} & $-8.7\%$ \cite{Fael:2020tow} & $-12\%$ \cite{Aquila:2005hq} &  $-1.8\%$ \cite{Fael:2020tow} & $-0.03\%$ \cite{Fael:2020tow} & $0.3\%$ \cite{Alberti:2013kxa} & -- & $0.14\%$ \\
        $\ell_1$ & $-0.5\%$ & $-1.1\%$ & $\mathcal{O}(0.5\%)$ \cite{Mannel:2010wj} & $\mathcal{O}(0.25)\%$ \cite{Mannel:2010wj} & $-0.2\%$ \cite{Gambino:2011cq} & $0.24\%$ \cite{Gambino:2011cq} & $-0.1\% $ \cite{Gambino:2011cq} & $-0.2\%$ \cite{Fael:2022frj} & $0.5\%$ \cite{Alberti:2013kxa} & -- & $0.01\%$ \\
        $\ell_2$ & $9.7\% $ & $-11.5\% $ & $\mathcal{O}(5\%)$ \cite{Mannel:2010wj} & $\mathcal{O}(1\%)$ \cite{Mannel:2010wj} & $-0.5\%$ \cite{Gambino:2011cq} & $1.7\%$ \cite{Gambino:2011cq} & $-0.6\%$ \cite{Gambino:2011cq} & $0.02\%$ \cite{Fael:2022frj} & $-2\%$ \cite{Alberti:2013kxa} & -- & $-0.03\%$ \\
        $\ell_3$ & $-255\%$ & $191\%$ & $\mathcal{O}(-30)\%$ \cite{Mannel:2010wj} & $\mathcal{O}(10\%)$ \cite{Mannel:2010wj} & $-3\%$ \cite{Gambino:2011cq} & $3.1\%$ \cite{Gambino:2011cq} & $-6\%$ \cite{Gambino:2011cq} & $1.4\%$ \cite{Fael:2022frj} & $-1.5\%$ \cite{Alberti:2013kxa} & -- & $0.03\%$ \\
        \hline
        $\left\langle m_X^2 \right\rangle$ & $-2.81\%$ & $5.3\%$ & $\mathcal{O}(-2\%)$ \cite{Mannel:2010wj} & $\mathcal{O}(-1\%)$ \cite{Mannel:2010wj} & $1.3\%$ \cite{Gambino:2011cq} & $-1.2\%$ \cite{Gambino:2011cq} & $0.6\% $ \cite{Gambino:2011cq} & $0.03\%$ \cite{Fael:2022frj} & $0.6\%$ \cite{Alberti:2013kxa} & -- & $-0.01\%$ \\
        $h_{2}$ & $788\%$ & $-628\%$ & $\mathcal{O}(100\%)$ \cite{Mannel:2010wj} & $\mathcal{O} (50\%) $ \cite{Mannel:2010wj} & $99\%$ \cite{Gambino:2011cq} & $85\%$ \cite{Gambino:2011cq} & $-34\%$ \cite{Gambino:2011cq} & $51\%$ \cite{Fael:2022frj} & $-143\%$ \cite{Alberti:2013kxa} & -- & $0.04\%$ \\
        $h_{3}$ & $-6633.5\%$ & $-22448.2\%$ & $\mathcal{O} (6000\%)$ \cite{Mannel:2010wj} & $\mathcal{O}(600\%)$ \cite{Mannel:2010wj} & $1761\%$ \cite{Gambino:2011cq} & $2020\%$ \cite{Gambino:2011cq} & $155\%$ \cite{Gambino:2011cq} & $-2000\%$ \cite{Fael:2022frj} & $-8475\%$ \cite{Alberti:2013kxa} & -- & $-0.2\%$ \\
        \hline
        $\left\langle q^2 \right\rangle$ & $-3.4\%$ & $-4\%$ & $0.6\%$ & $-$ & $1.4\%$ \cite{Mannel:2021zzr} & -- & -- & $-0.5\%$ \cite{Fael:2022frj} & $-0.7\%$ \cite{Mannel:2021zzr} & $-1.8\%$ \cite{Mannel:2021zzr} & $0.06\%$ \\
        $q^2_{\mathrm{central},2}$ & $-14.4\%$ & $-23.5\%$ & $4\%$ & $-$ & $2.8\%$ \cite{Mannel:2021zzr} & -- & -- & $-1.1\%$ \cite{Fael:2022frj} & $-2.7\%$ \cite{Mannel:2021zzr} & $-9.8\%$ \cite{Mannel:2021zzr} & $0.04\%$ \\
        $q^2_{\mathrm{central},3}$ & $-32.2\%$ & $-87.2\%$ & $18\%$ & $-$ & $-5.8\%$ \cite{Mannel:2021zzr} & -- & -- & $-1.9\%$ \cite{Fael:2022frj} & $-3.8\%$ \cite{Mannel:2021zzr} & $-33\%$ \cite{Mannel:2021zzr} & $-0.2\%$ \\
        $q^2_{\mathrm{central},4}$ & $-31.9\%$ & $-71.1\%$ & $16\%$ & $-$ & $4.8\%$ \cite{Mannel:2021zzr} & -- & -- & $-2.4\%$ \cite{Fael:2022frj} & $-5.5\%$ \cite{Mannel:2021zzr} & $-27\%$ \cite{Mannel:2021zzr} & $0.02\%$ \\
        \hline
    \end{tabular}
    }
\end{sidewaystable*}

\section{Cut dependence} \label{sec:CutDependence}
\begin{figure*}[!ht]
    \centering
    \includegraphics[width=0.9\textwidth]{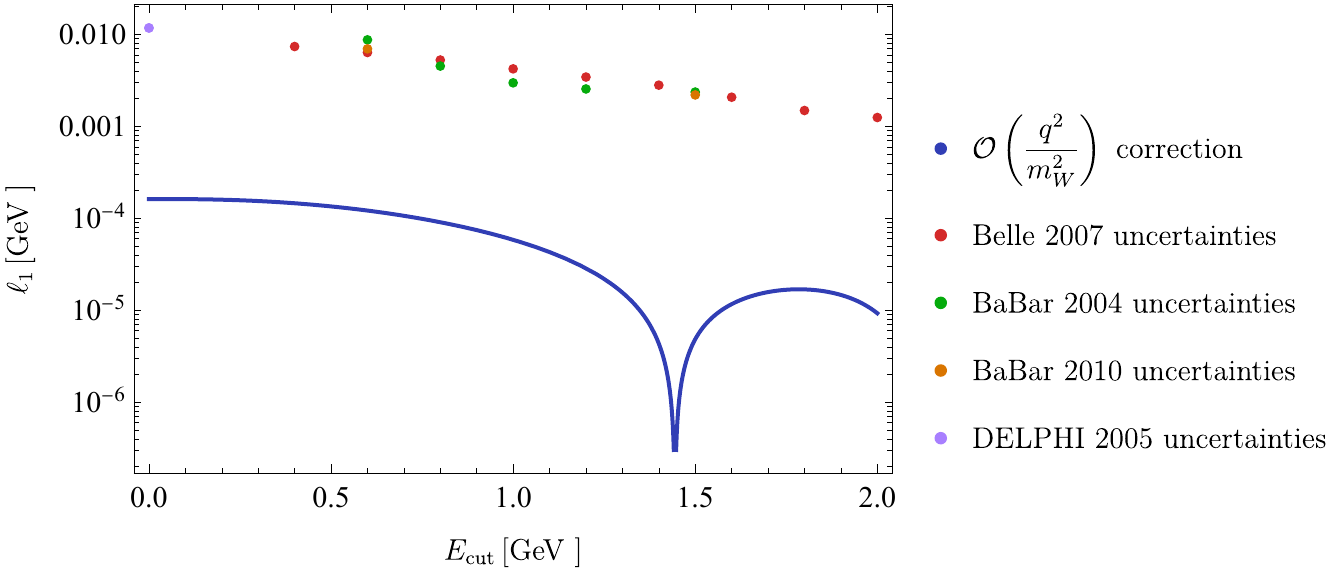}
    \caption{Cut dependence of the first lepton energy moment}
    \label{fig:FirstLeptonMomentCutDependence}
\end{figure*}

\begin{figure*}[!ht]
    \centering
    \includegraphics[width=0.9\textwidth]{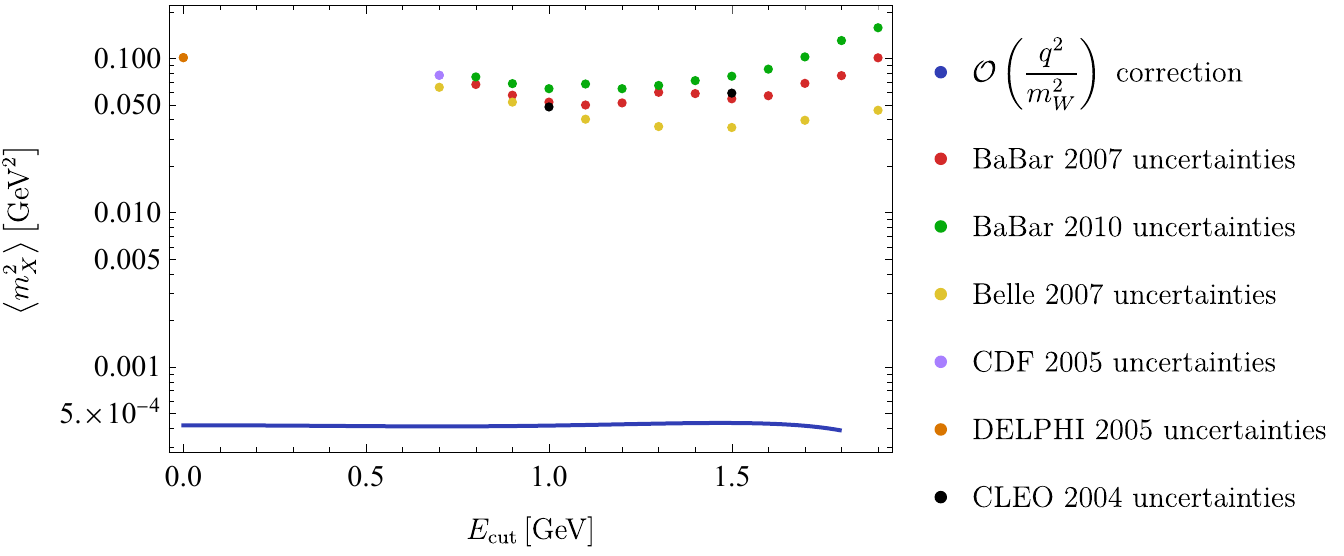}
    \caption{Cut dependence of the first hadronic mass moment}
    \label{fig:mx2MomentCutDependence}
\end{figure*}

\begin{figure*}[!ht]
    \centering
    \includegraphics[width=0.9\textwidth]{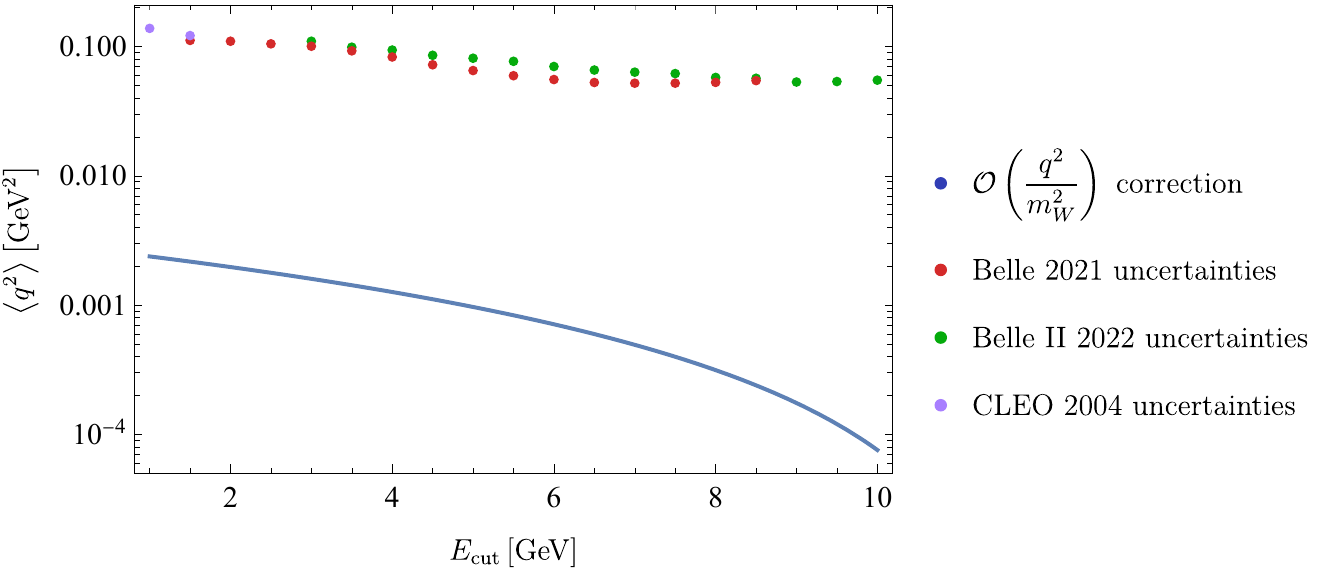}
    \caption{Cut dependence of the first leptonic invariant mass moment}
    \label{fig:q2MomentCutDependence}
\end{figure*}

In Figures \ref{fig:FirstLeptonMomentCutDependence}-\ref{fig:q2MomentCutDependence} we present the dependencies of the moments on the phase space cuts applied in experiments.

\end{appendices}

\clearpage


\bibliography{bibliography}


\end{document}